\documentclass[a4paper,fleqn]{cas-dc}

\usepackage[numbers,sort&compress]{natbib}

\RequirePackage{color}

\begin{document}
\let\WriteBookmarks\relax
\def\floatpagepagefraction{1}
\def\textpagefraction{.001}

\shorttitle{Average scattering entropy for periodic, aperiodic and
  random distribution of vertices in simple quantum graphs}

\shortauthors{A. A. Silva et~al.}

\title [mode = title]
{Average scattering entropy for periodic, aperiodic and
  random distribution of vertices in simple quantum graphs}

\author[1]{Alison A. Silva}[orcid=0000-0003-3552-8780]

\affiliation[1]{organization={Universidade Estadual de Ponta Grossa},
            addressline={Programa de P\'os-Gradua\c{c}\~ao},
            city={Ponta Grossa},
            citysep={}, 
            postcode={84030-900},
            state={Paran\'a},
            country={Brazil}}

\author[1,2]{Fabiano M. Andrade}[orcid=0000-0001-5383-6168]
\ead{fmandrade@uepg.br}
\affiliation[2]{organization={Universidade Estadual de Ponta Grossa},
            addressline={Departamento de Matem\'atica e Estat\'istica},
            city={Ponta Grossa},
            citysep={}, 
            postcode={84030-900},
            state={Paran\'a},
            country={Brazil}}

          \cormark[1]
\cortext[cor2]{Corresponding author}

\author[3]{D. Bazeia}[orcid=0000-0003-1335-3705]
\affiliation[3]{organization={Universidade Federal da Para\'iba},
            addressline={Departamento de F\'isica},
            city={Jo\~{a}o Pessoa},
            citysep={}, 
            postcode={58051-900},
            state={Para\'iba},
            country={Brazil}}

\begin{abstract}
This work deals with the average scattering entropy of quantum graphs.
We explore this concept in several distinct scenarios that involve
periodic, aperiodic and random distribution of vertices of distinct
degrees.
In particular, we compare distinct situations to see how they behave as
we change the arrangements of vertices and the topology of the proposed
structures.
The results show that the average scattering entropy may depend
on the number of vertices, and on the topological
disposition of vertices and edges of the quantum graph.
In this sense, it can be seen as another tool to be used to explore
topological effects of current interest for quantum systems.
\end{abstract}



\begin{keywords}
  Entropy \sep Scattering \sep Quantum graphs \sep Topology
\end{keywords}

\maketitle

\section{Introduction}

The main aim of the present work is to study specific properties of
quantum graphs \cite{Book.Berkolaiko.2012}.
In particular, we will investigate the average scattering entropy (ASE),
a concept based on the Shannon entropy \cite{Book.Shannon.1963} which
was recently introduced in Ref. \cite{PRA.103.062208.2021}, that allows
us to associate a global quantity to a given quantum graph (QG).
The ASE is calculated from the scattering properties of a QG,
so we take advantage of the scattering Green's function approach for QG
developed in Refs. \cite{PRA.98.062107.2018,PR.647.1.2016} to implement
the investigation.
Here we focus on the  calculation of this quantity in several distinct
situations, to show that the procedure works adequately and also, to see
how it changes as we modify the array of vertices and leads following
some general possibilities, as the periodic, aperiodic and random
distribution of vertices in the graphs.
The interest is to add more information concerning the behavior
of the ASE as we change the way the vertices dispose themselves in each
QG.
Besides, we want to add more motivation with the study of QG
that may be directly connected to the study of quantum walks
\cite{JPA.37.6675.2004,Incollection.2006.Tanner,PRA.48.1687.1993,
arXiv:quant-ph.0010117,CP.44.307.2003}, quantum walks in optical
lattices \cite{PRA.66.052319.2002}, and networks of soft active matter
\cite{RMP.85.1143.2013,NRM.2.17048.2017,PRL.9.098002.2020}.

\begin{figure}[b]
  \centering
  \includegraphics[width=0.9\columnwidth]{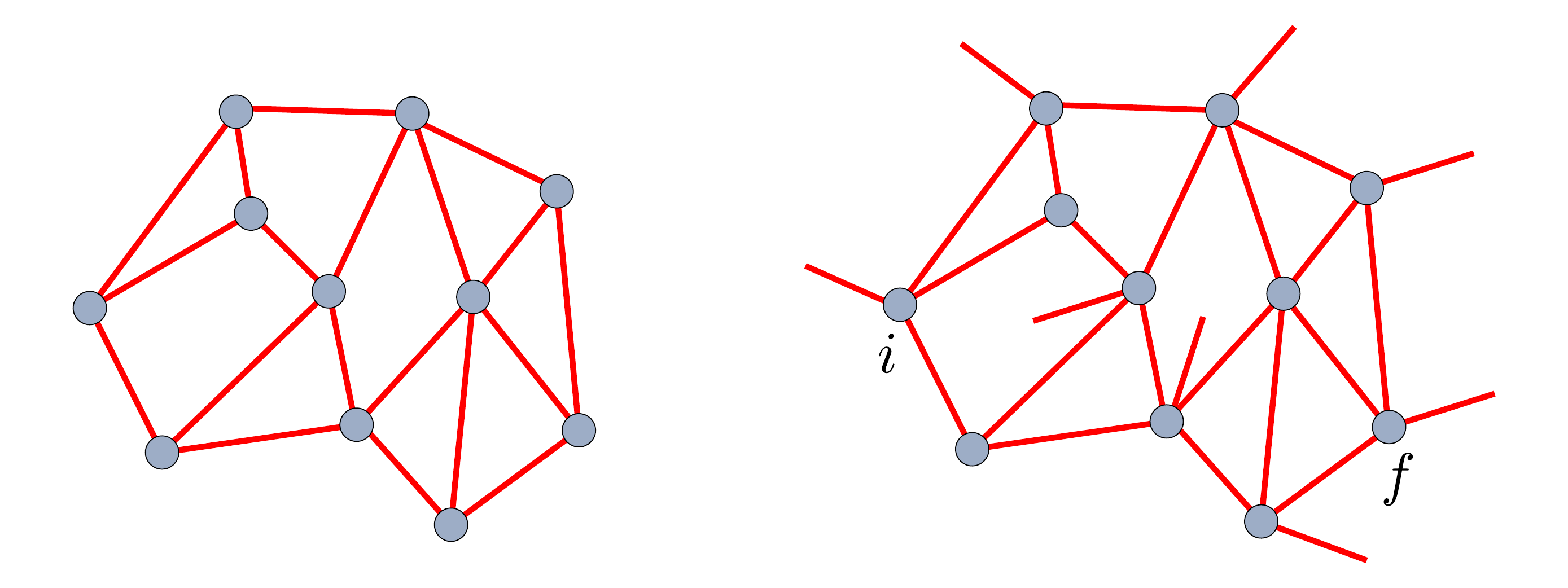}
  \caption{(Color online)
    Closed quantum graph with $11$ vertices and $19$ edges
    (left) and the associated open quantum graph with $8$
    leads added (right), with $i$ and $f$ identifying the entrance and
    exit scattering channels.
  }
  \label{fig:fig1}
\end{figure}

\begin{figure}
  \centering
  \includegraphics[width=0.9\columnwidth]{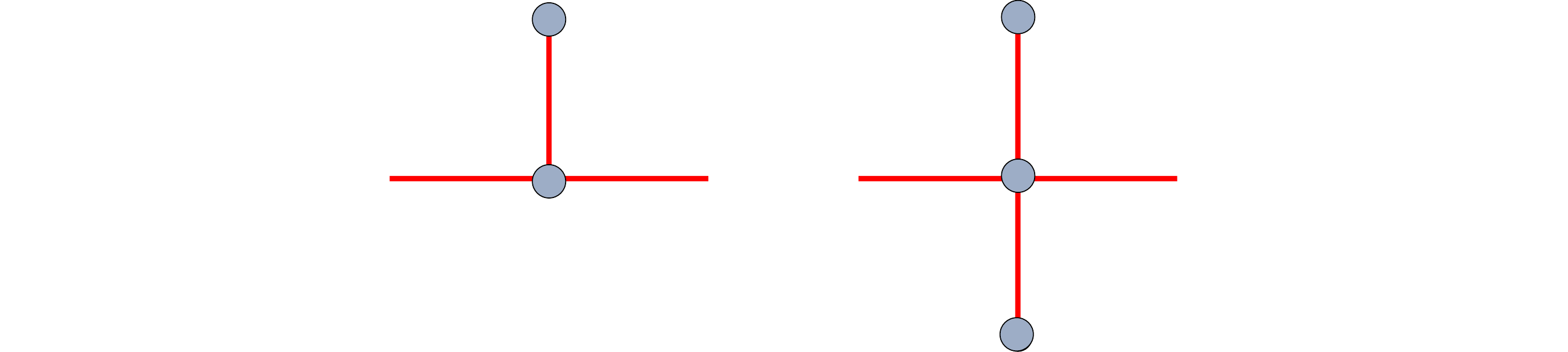}
  \caption{(Color online)
    Two very simple quantum graphs with two external leads.
    They are the $\alpha$ (left) and $\beta$ (right) structures, and
    have a single vertex of degree $3$ (left) or $4$ (right) attached to
    one or two dead end vertices of degree $1$, respectively.
  }
  \label{fig:fig2}
\end{figure}

As one knows, a QG is a structure associated with
an arbitrary arrangement of vertices, edges and leads, and in
Fig. \ref{fig:fig1} we depict two quantum graphs, both having $11$
vertices and $19$ edges, but in the structure on the right
one attaches $8$ leads, as appropriate for the study of scattering
properties of quantum graphs.
Beyond paying closer attention to the periodic, aperiodic and random
distribution of vertices in the QG, we will also deal with
topology and geometry, to see how they can contribute to change the
ASE contents of each QG.
Although the periodic array of vertices in a QG
is of direct interest to physics, the possibility of considering
aperiodic and random distribution of vertices is also of interest,
although more involved.
This is related to the possibility to select two
very simple but distinct structures and use them to build distinct
arrangements of vertices and edges.
We name these elementary structures as $\alpha$ and $\beta$, and
depict their forms in Fig. \ref{fig:fig2}.
These QGs  were already studied in
\cite{PRA.103.062208.2021} and despite their simplicity, they engender
interesting scattering properties and the respective ASE values are
calculated at the end of Sec. \ref{sec:sqg}

In the investigation developed in Ref. \cite{PRA.103.062208.2021}, we
have introduced the ASE owing to associate a global value connected to
the complicated energy-dependent scattering probabilities of a given
quantum graph, based on the Shannon entropy.
As one knows, for a system having a set of $n$ equally probable
states, the Shannon entropy gives $\log_2 n$, so it increases as the
number of states increases. As a consequence, a fair die has more
information value than a fair coin. In the case of the ASE, we are
dealing with the scattering probabilities of open quantum graphs (see
Sec. \ref{sec:sqg} for details), but its interpretation is similar to
the interpretation of the Shannon entropy.

In this work, we use the $\alpha$ and $\beta$ structures to build
periodic, aperiodic and random arrangements on the line and on
the circle and study how the ASE changes as we add more and more
vertices to the QGs.
A motivation of current interest is that $\alpha$ and $\beta$ can be
seen as a quantity having two distinct conformations as in a two-state
quantum system.
This possibility opens an interesting route of investigation, which is
directly related to the study of quantum walks using neutral atoms
trapped in optical lattices, for which the investigations
\cite{NJP.16.123052.2014,PRA.66.052319.2002} may be considered for
applications of practical interest.
To make the investigation more general, we also study other
arrangements, in particular the two-dimensional case where the network
conforms itself as flat and curved defect-like structure, which is
directly related to the model investigated in
Ref. \cite{PRL.9.098002.2020},  that is of interest to the study of
pattern formation in systems of active soft matter
\cite{RMP.85.1143.2013}.

We also study other networks which require working in two and three
dimensions.
Examples are the sequences of squares which can be disposed as flat
stripe-like configurations in two dimensions (2D), and the regular
triangular prisms that form triangular tube-like configurations in three
dimensions (3D).
These arrangements will also be investigated in this work, to see how
they contribute to change the corresponding ASE values, as we increase
the number of elements in 2D and 3D networks.

In addition to the direct interest related to thermodynamics,
statistical and quantum information, there are several other
possibilities connected to the subject of the present work.
In high energy physics, for instance, the concept of configurational
entropy introduced in Ref. \cite{PLB.713.304.2012}, which is also based
on the Shannon entropy, has been used in several distinct directions to
analyse stability of physical systems that are localized in space:
in the investigation of mesons propagating in chiral and gluon
condensates in a gravity background \cite{PRD.98.126011.2018}, in the
examination of heavy mesons in the quark gluon plasma
\cite{PLB.787.16.2018,PLB.811.135918.2020}, in the study of the magnetic
structure of skyrmions in planar magnetic materials
\cite{JMMM.475.734.2019,PLA.392.127170.2021}, in the investigation of the
soft wall background as a function of the temperature
\cite{PLB.820.136485.2021}, the mass spectra of meson resonances
\cite{PRD.103.106027.2021} and anti-de Sitter graviton stars
\cite{PLB.823.136729.2021}, to quote just a few recent works on the
subject.

We organize the present investigation as follows:
in the Sec. \ref{sec:sqg} we review the scattering in QGs using the
Green's function approach, and the concept concerning the ASE on general
grounds.
After briefly describing the formalism, we then include in Sec.
\ref{sec:specific} the main results of the work, calculating the ASE for
several open and closed arrangements of vertices in distinct scenarios,
involving periodic, aperiodic and random distributions of elementary
structures, and other two
and three dimensional arrangements of vertices, edges and external
leads.
We then end the work in Sec. \ref{sec:end}, where we add some comments
and suggest new lines of research of current interest related to the
subject of study in the present investigation.

\section{Scattering in quantum graphs}
\label{sec:sqg}

As one knows, a QG can be described as a triple
$\{\Gamma(V,E), H, \text{BC}\}$, consisting of a metric graph
$\Gamma(V,E)$, a differential operator $H$ and a set of boundary
conditions, $\text{BC}$ \cite{Book.Berkolaiko.2012}.
A metric graph $\Gamma(V,E)$ is a set of $v$ vertices,
$V=\{1,\ldots,v\}$, and a set of $e$ edges, $E=\{e_1,\ldots,e_e\}$,
where each edge links a pair of vertices $e_s=\{i,j\}$, with
non-negative lengths $\ell_{e_{s}}\in (0,\infty)$.
Here we consider the free Schr\"odinger operator
$H=-(\hbar^2/2m)d^2/dx^2$ acting on each edge and the most natural set
of  boundary conditions on the vertices, namely, the Neumann boundary
conditions.
The graph topology is defined by its adjacency matrix $A(\Gamma)$ of
dimension $v \times v$; their elements $A_{ij}(\Gamma)$
are $1$ if the vertices $i$ and $j$ are connected and $0$ otherwise.
We introduce an open QG, $\Gamma^{l}$, which is suitable to
study scattering problems, by adding $l$ leads (semi-infinite edges) to
its vertices, as illustrated in  Fig.
\ref{fig:fig1} (right).
The open QG $\Gamma^{l}$ can then be used to describe a scattering
system with $l$ scattering channels which is characterized by the
energy-dependent global scattering matrix
$\boldsymbol{\sigma}_{\Gamma^{l}}(k)$, where $k$ is the wave number,
which is related to the energy by the standard expression
$E=\hbar^2 k^2/2m$, and the matrix elements are given by the scattering
amplitudes $\sigma_{\Gamma^{l}}^{(f,i)}(k)$, where $i$ and $f$
represents the entrance and exit scattering channels, respectively.

\subsection{Scattering Amplitudes}

The calculation of the scattering amplitudes
$\sigma_{\Gamma^{l}}^{(f,i)}(k)$ is based on the scattering Green's
function approach developed in Refs.
\cite{PRA.98.062107.2018,PR.647.1.2016}.
This procedure was recently used to study narrow peaks of full
transmission and transport in simple quantum graphs in
Refs. \cite{PRA.100.62117.2019,EPJP.135.451.2020}, which has inspired us
to introduce the ASE in Ref. \cite{PRA.103.062208.2021} and explore it
in the present study.
In this manner, the scattering amplitudes are given by
\begin{equation}
  \label{eq:Srs}
  \sigma_{\Gamma^{l}}^{(f,i)}(k) =
  \delta_{fi}r_{i}
  +
  \sum_{j \in E_{i}}  A_{ij} P_{ij}^{(f)}(k)t_{i},
\end{equation}
where $i$ and $f$ are the entrance and exit scattering channels,
$E_i$ is the set of neighbor vertices connected to $i$, and $r_i$ and
$t_i$ are the reflection and transmission amplitudes at the vertex $i$,
which for Neumann boundary conditions (Neumann vertices), are given by
\cite{AP.55.527.2006}
\begin{equation}
  \label{eq:scatt_amp}
  r_{i} = \frac{2}{d_{i}} -1, \qquad
  t_{i} = \frac{2}{d_i},
\end{equation}
where $d_i\geq 2$ is the degree of the vertex $i$ (the total number of
edges and/or leads attached to it).
Neumann vertices of degree one have $r_i=1$.
The quantities $P_{ij}^{(f)}(k)$ are the families of paths between the
vertices $i$ and $j$, and they are given by
\begin{equation}
  \label{eq:pij}
  P_{ij}^{(f)}(k)
  =
   z_{ij}\delta_{fj} t_{j}
  +
  z_{ij}  P_{ji}^{(f)}(k) r_{j}
  +z_{ij} \sum_{l \in {E_{j}^{i,f}}}  A_{jl} P_{jl}^{(f)}(k) t_{j},
\end{equation}
where $z_{ij}= e^{i k \ell_{s}}$ with $\ell_{s}$ representing the length
of the edge $e_{s}=\{i,j\}$ connecting $i$ and $j$; also, $E_{j}^{i,f}$
stands for the set of neighbors vertices of $j$ but with the vertices
$i$ and $f$ excluded.
The family $P_{ji}^{(f)}(k)$ is obtained from the above equation by
swapping $i \leftrightarrow j$ and the number of family of paths is
always twice the number of edges of the graph under investigation.
The family of paths altogether form an inhomogeneous system of equations
and its solution leads to the scattering amplitude
$\sigma_{\Gamma^{l}}^{(f,i)}(k)$ \cite{PRA.98.062107.2018}.

\subsection{Average Scattering Entropy}
\label{sec:ase}

Consider the scattering on a QG $\Gamma^{l}$ as described above.
By fixing the entrance channel, say $i$, this scattering system is
characterized by $l$ quantum amplitudes, which are given by Eq.
\eqref{eq:Srs}.
This defines a set of $l$ scattering probabilities
\begin{equation}\label{proba}
  p_{\sigma_{\Gamma^{l}}}^{(j)}(k)=|\sigma_{\Gamma^{l}}^{(j,i)}(k)|^{2},
\end{equation}
for a quantum particle entering the graph with wave number $k$ through
the fixed vertex $i$, and exiting the graph by some other vertex $j$
(also including the vertex $i$ itself).
These probabilities fulfills the constrain
\begin{equation}
  \sum_{j=1}^{l}p_{\sigma_{\Gamma^{l}}}^{(j)}(k)=1,
\end{equation}
which is a consequence of the probability conservation in the scattering
process, or better, of the unitarity of the scattering matrix.
Thus, the scattering process that occurs in a QG is analogous to a
random variable with $l$ possible outcomes.
With this in mind, for graphs where their scattering
probabilities are periodic, in Ref.
\cite{PRA.103.062208.2021} we
introduced the ASE which is given by
\begin{equation}
  \label{eq:AS}
  \bar{H}(\sigma_{\Gamma^{l}})=\frac{1}{K}
  \int_0^K H_{\sigma_{\Gamma^{l}}} (k)\, dk,
\end{equation}
where $K$ is the period of the scattering probability, and
\begin{equation}
  \label{eq:S_perk}
  H_{\sigma_{\Gamma^{l}}} (k)=
  -\sum_{j=1}^{l} p_{\sigma_{\Gamma^{l}}}^{(j)}(k)
  \log_2 p_{\sigma_{\Gamma^{l}}}^{(j)}(k),
\end{equation}
is the Shannon entropy which encodes the informational content of the
scattering process on a QG as a function of $k$.
Therefore, the ASE encodes all the complicated behavior of the
scattering probabilities of a QG into a single number.
It is interesting to observe that when the scattering probability for
all the $l$ scattering channels is equal to
$p_{\sigma_{\Gamma^{l}}}^{(j)}(k)=1/l$, $H_{\sigma_{\Gamma^{l}}} (k)$
assumes its maximum value $\log_2l$.
On the other hand, the minimum occurs when all the scattering
probabilities are $0$ but one is $1$, which corresponds to
a full reflection or a full transmission.

Evidently, the probability displayed in Eq. \eqref{proba} also depends
on the incoming channel, and so does the ASE in
Eq. \eqref{eq:AS} and the entropy in Eq. \eqref{eq:S_perk}.
Moreover, we notice here that associating the entries of a unitary
matrix with transition probabilities has also been done in
Ref. \cite{JPA.34.8485.2001} in the context of stochastic dynamics on a
graph, proposing to study unitary matrix ensembles defined in terms of
unitary stochastic transition matrices associated with Markov processes
on graphs, which may motivate another line of study.
We also notice that the above procedure can be directly used to
calculate the ASE values for the two simple graphs displayed in
Fig. \ref{fig:fig2}: the $\alpha$ and $\beta$ structures give $0.503258$
and $0.557305$, respectively.
We can verify that these results are independent of the boundary
conditions for the dead end vertices of degree $1$ being Dirichlet or
Neumann; see, e.g., Ref. \cite{PRA.103.062208.2021} for other details on
this issue.

\section{Results}
\label{sec:specific}

Let us now investigate some specific QG configurations to see how the
ASE behaves as we change the number of vertices and the corresponding
degrees, and the way they are assembled on the line and on the
circle, with periodic, aperiodic and random dispositions, and in other
situations having geometric and spatial modifications.
The several possibilities investigated are described in the subsections
that follows.

\subsection{Periodic arrangements}
\label{sec:regular}

\begin{figure}[t]
  \centering
  \includegraphics[width=0.9\columnwidth]{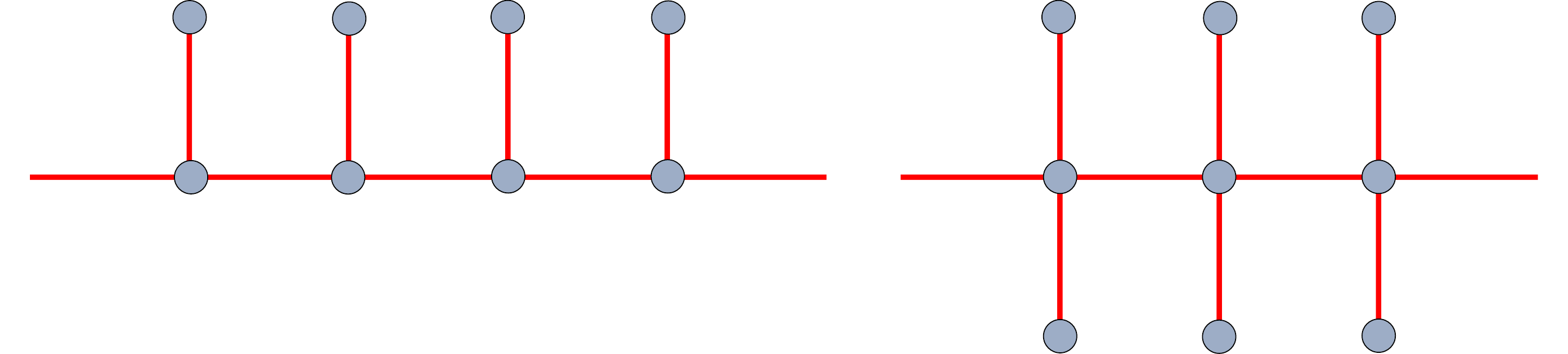}
  \caption{(Color online)
    The $\alpha_4$ (left) and $\beta_3$ (right) arrangements which we
    use to illustrate the general case on the line.}
  \label{fig:fig3}
\end{figure}

\begin{figure}[b]
  \centering
  \includegraphics[width=0.9\columnwidth]{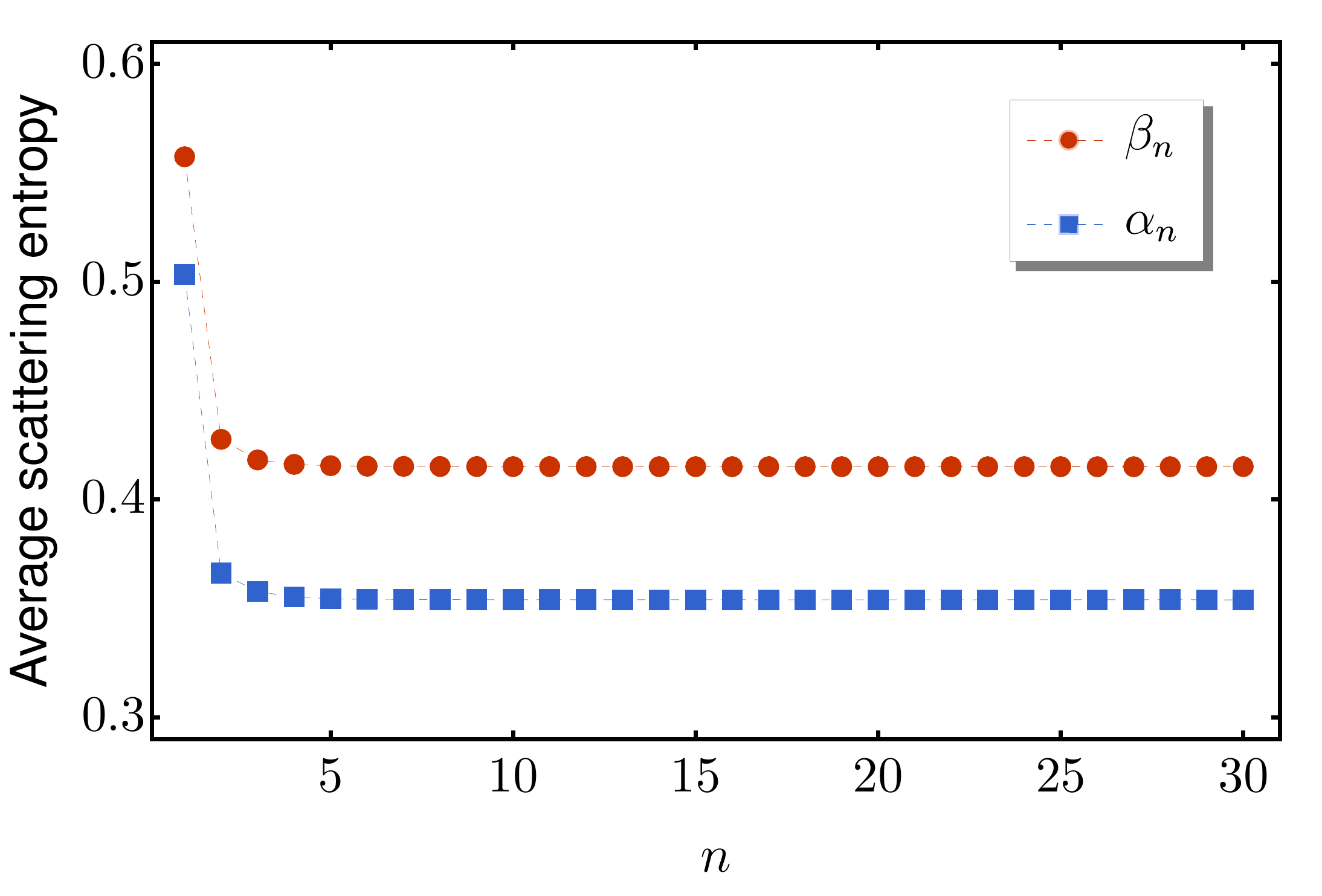}
  \caption{(Color online)
   Average scattering entropy of the quantum graphs $\alpha_n$  (blue)
   and $\beta_n$ (red) on the line, for several values of $n$.
  }
  \label{fig:fig4}
\end{figure}

We start using a regular distributions of vertices of degree $d=3$ and
$d=4$ on the line, generically represented by $\alpha_n$ and $\beta_n$,
with $\alpha$ and $\beta$ already introduced in Fig. \ref{fig:fig2}, and
with $n=1,2,3,\cdots$ standing for the number of replications in the
arrangements.
We are considering ideal vertices, and we use Neumann boundary
conditions for the dead end vertices of degree $1$, wherever they appear
in the quantum graphs.
Also, in these arrangements, all the edges connecting two vertices have
the same length, i.e., we are considering \textit{equilateral} QGs.
An illustration is depicted in Fig. \ref{fig:fig3} for the cases
$\alpha_4$ and $\beta_3$.
We take advantage of the results obtained in
Ref. \cite{PRA.103.062208.2021} and display in red and blue in
Fig. \ref{fig:fig4} the several values of the ASE with $\alpha$  (blue)
and $\beta$ (red) vertices on the line.
We notice that the value rapidly diminishes and saturates to a constant
as $n$, the number of replications, increases to larger and larger
values.
Also, it is always higher for the arrangements with $\beta$ vertices.
In this sense, one sees that although the degree $d$ is significant, the
number of replications or the size of the lattice seems to play no
important role for $n$ greater than $3$.
This means that the ASE is importantly affected by the degree of the
vertices, although it is practically insensitive to the addition of
extra vertices in chains of three or more vertices.
However, it varies significantly when one changes from one to two
replications, meaning that small structures feel the ASE more
importantly.

\begin{figure}[t]
  \centering
  \includegraphics[width=0.9\columnwidth]{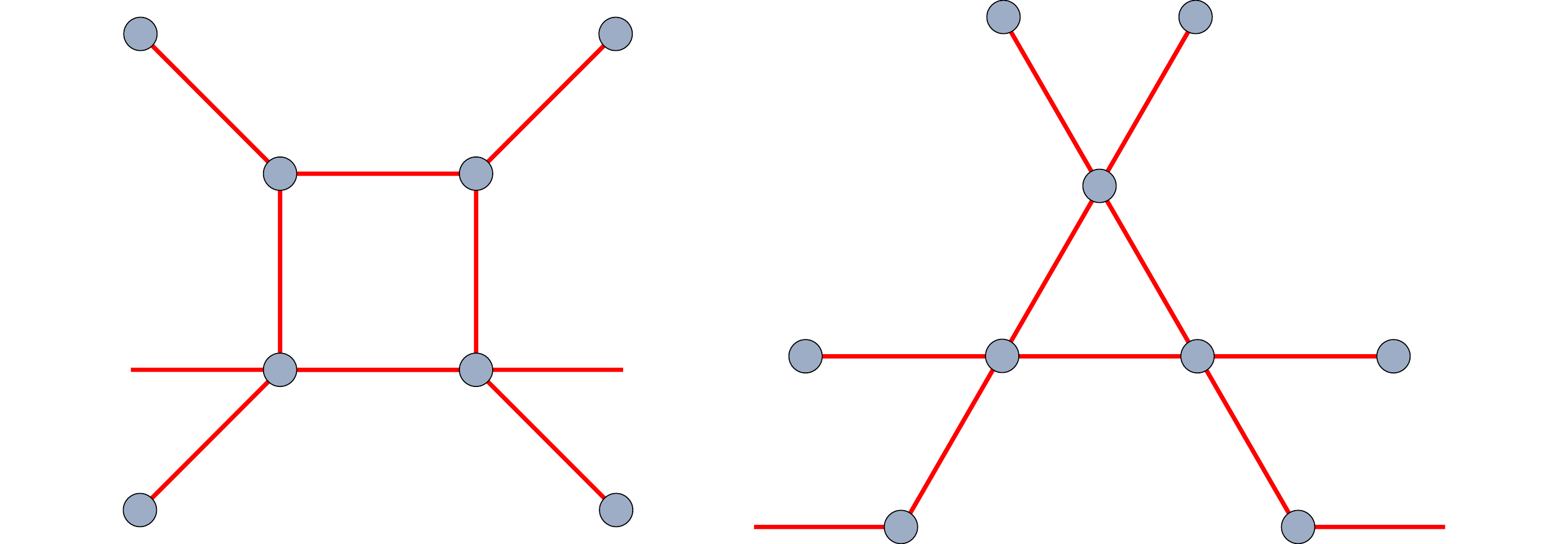}
  \caption{(Color online)
    Two quantum graphs with $\alpha_4$ (left) and $\beta_3$
    (right), which illustrate the general case of periodic
    structures on the circle.
  }
  \label{fig:fig5}
\end{figure}

Another situation of interest concerns periodic arrangement of vertices
on the circle, and here we study the case with $\alpha$ and $\beta$
vertices, and compare the results with the results already obtained on
the line.
In Fig. \ref{fig:fig5} we display two QGs to illustrate the
general situation in the two cases on the circle.
We notice that all the degree 1 vertices are attached to vertices of
degree three for the $\alpha$ configurations, and four, for the $\beta$
configurations.
On the circle we then have ring-like structures formed by vertices of
degree 3 and 4, for the $\alpha$, and $\beta$ structures, respectively,
and we then have several different ways to probe them under scattering
adding two external leads.
Here we will consider the following two cases, which we believe are of
interest for the transport in molecules in break
junctions \cite{NRP.1.211.2019,NRP.1.381.2019} and in microwave networks
\cite{PRE.69.056205.2004}, for instance.
The first case refers to the addition of two external leads, one at one
of the vertices in the ring-like structure, and the other one at the
neighbor vertex in the ring.
This is illustrated in Fig. \ref{fig:fig5}, left, with the two
external leads, and we identify this type of arrangement as Circle. In the
second case, we also attach two external leads, one to a vertex of
degree 1 which is connected to a vertex in the ring, and the other to a
vertex of degree 1 which is connected to the neighbor vertex in the
ring.
This is also illustrated in Fig. \ref{fig:fig5}, right, with the two
external leads, and we identify these configurations as Circle2.
We notice that in the first case, we keep the total number of vertices
and edges, but we increase by 1 the degree of two vertices in the ring;
and in the second case, we remove 2 vertices of degree 1 and two edges,
since vertices of degree 2 are transparent, but we keep the degree
of the vertices in the ring.

\begin{figure}[b]
  \centering
  \includegraphics[width=0.9\columnwidth]{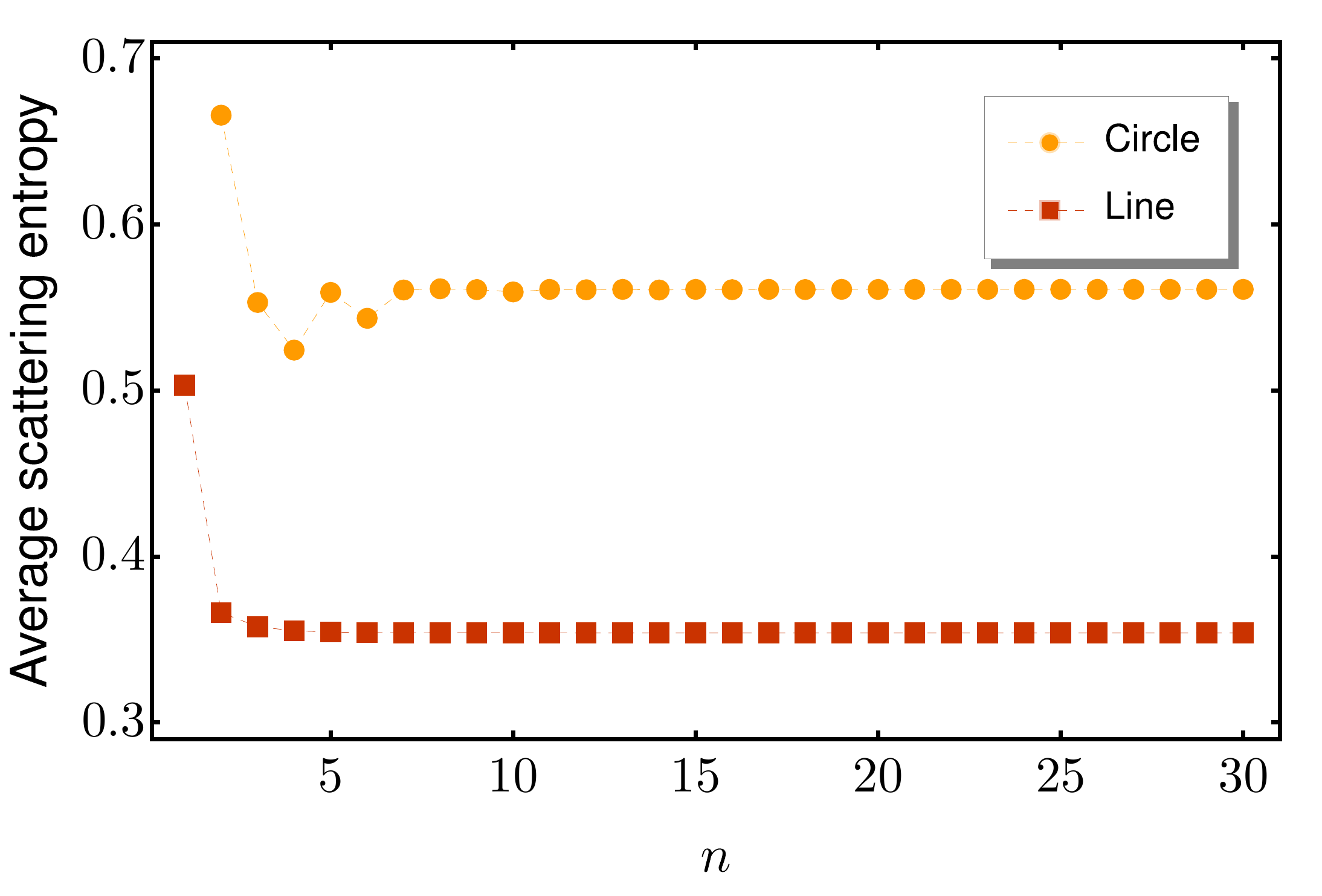}
  \caption{(Color online)
    Average scattering entropy of quantum graphs formed by $\alpha$
    vertices of degree $3$ on the line (red) and on the circle
    (orange).
  }
  \label{fig:fig6}
\end{figure}

\begin{figure}[t]
  \centering
  \includegraphics[width=0.9\columnwidth]{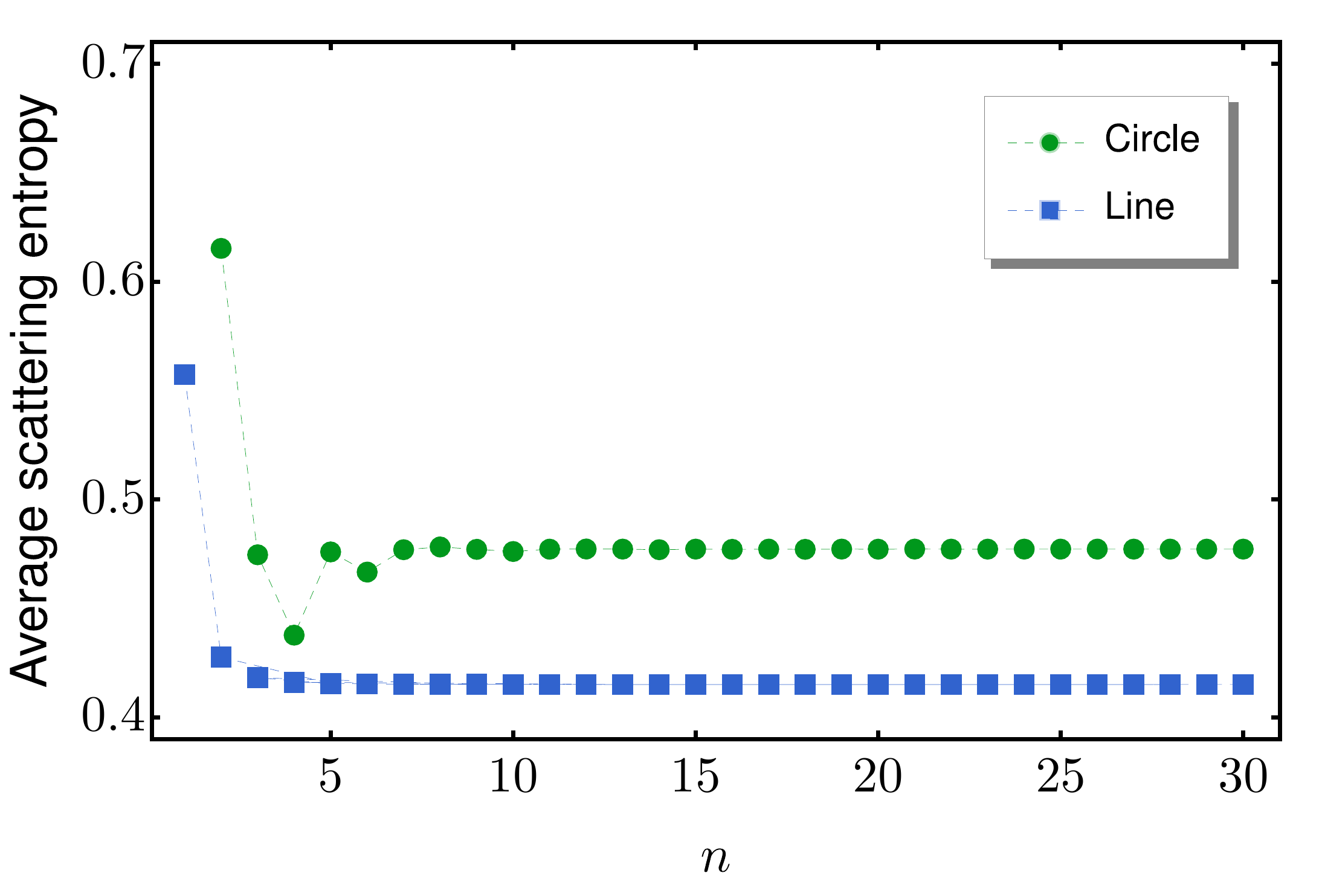}
  \caption{(Color online)
    Average scattering entropy of quantum graphs formed by $\beta$
    vertices of degree $4$ on the line (blue) and on the circle
    (green).
  }
  \label{fig:fig7}
\end{figure}

\begin{figure}[b]
  \centering
  \includegraphics[width=0.9\columnwidth]{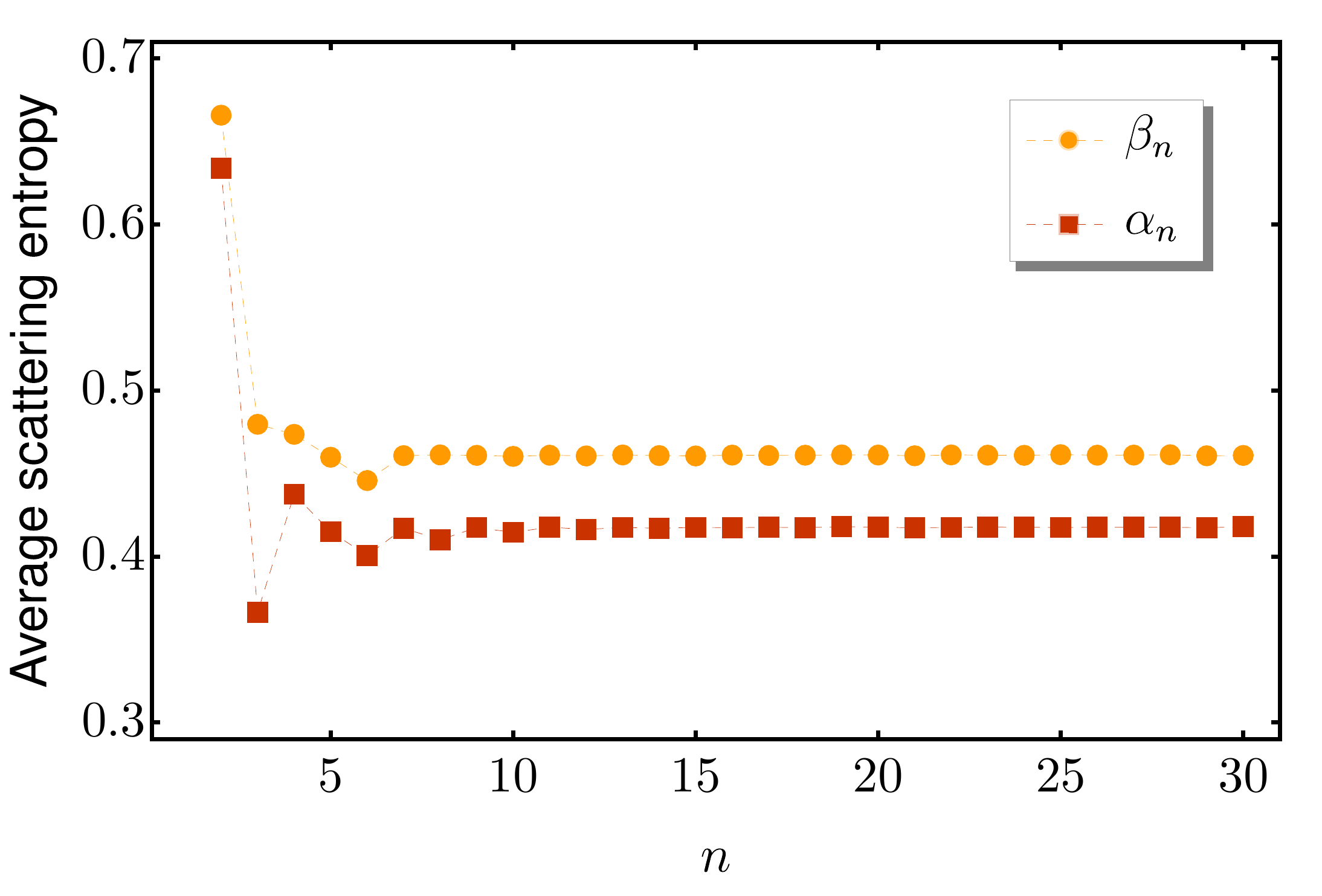}
  \caption{(Color online)
    Average scattering entropy of the quantum graphs $\alpha_n$ (red)
   and $\beta_n$ (yellow) for configurations of the Circle2 type, for
   several values on $n$.
  }
  \label{fig:fig8}
\end{figure}

Now, we follow the approach in Ref. \cite{PRA.103.062208.2021} and
depict the results in Figs. \ref{fig:fig6} and \ref{fig:fig7}
on the line and on the circle in the first case.
We notice that the values on the circle are always higher than on the
line.
We also notice from Figs. \ref{fig:fig6} and \ref{fig:fig7} that for
$n\leq7$, the ASE varies more significantly on the circle than on the
line. We also consider the ASE for configurations of the Circle2 type.
The results appear in Fig. \ref{fig:fig8}, and we notice that the
results for $\beta$ configurations are always higher than in the case of
the $\alpha$ type.

\subsection{Aperiodic arrangements}
\label{sec:fibonacci}

Let us now consider QG configurations with aperiodic compositions of
$\alpha$ and $\beta$ vertices, on the line and on the circle.
In order to implement the aperiodic disposition, we follow
\cite{PRL.93.190503.2004} and choose $\alpha$ and $\beta$ with the
following rules: $\alpha\to\alpha\beta$ and $\beta\to \alpha$; one
starts with $\alpha$ and implement the rules, to get the
sequences: $\alpha$, $\alpha\beta$, $\alpha\beta\alpha$,
$\alpha\beta\alpha\alpha\beta$, etc.
This reproduces the Fibonacci sequence of numbers, so using
the $\alpha$ and $\beta$ vertices, one can then construct aperiodic
sequences of vertices on the line and on the circle.
In Ref. \cite{PRL.93.190503.2004} one can find interesting physical
motivation and experimental implementation for these aperiodic
arrangements of vertices.
The fact that $\alpha$ and $\beta$ correspond to vertices of
two different degrees, it allows us to associate them to two different
states of each individual vertex.
This can be used to make a direct connection between arrangements of
vertices and optical lattices of Rubidium neutral atoms, for instance,
since Rubidium may be treated as a two-state system; see, e.g., Ref. \cite{PRA.66.052319.2002}.

\begin{figure}[t]
  \centering
  \includegraphics[width=0.9\columnwidth]{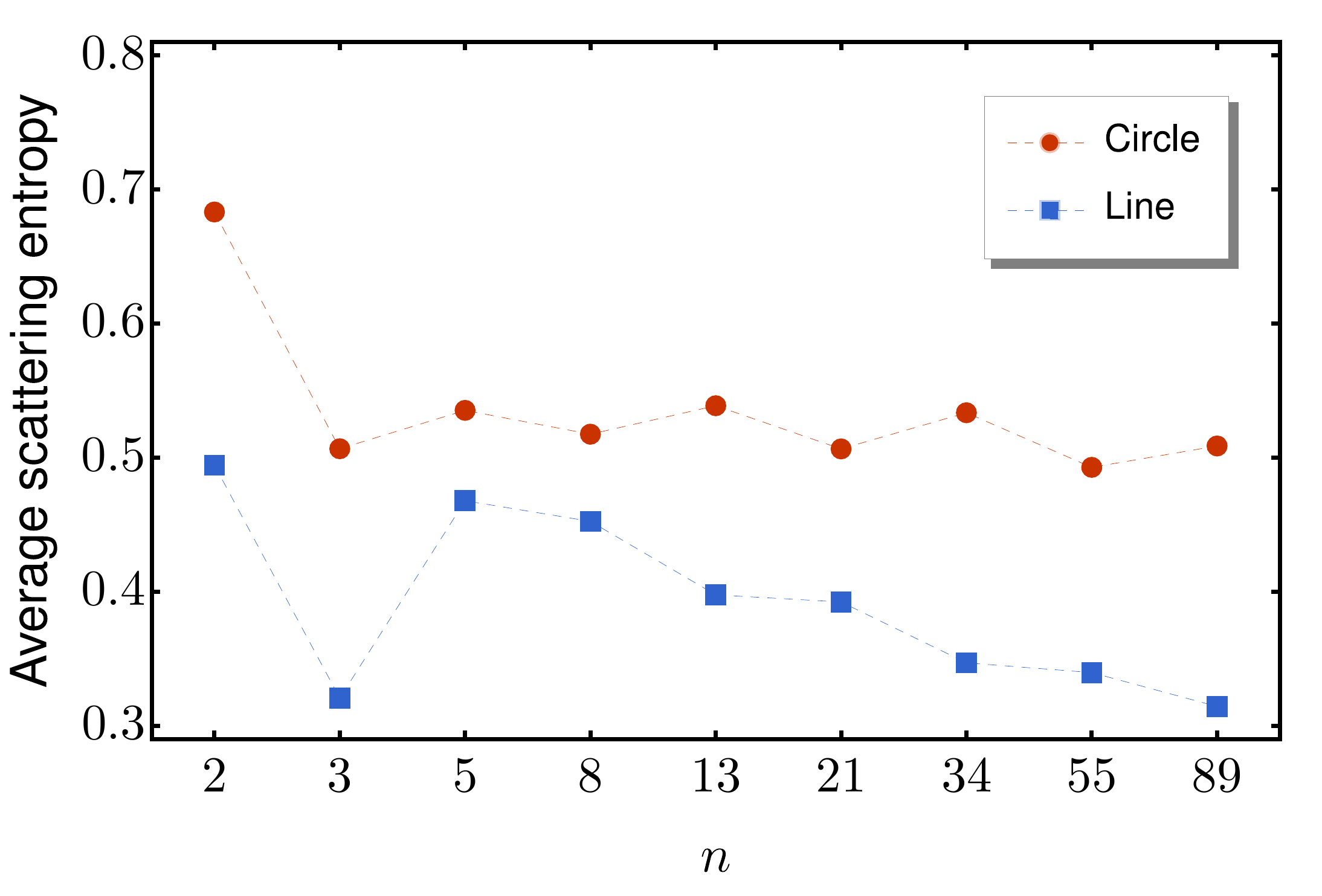}
  \caption{(Color online)
   Average scattering entropy of quantum graphs on the line (blue) and
   on the circle (red) identified by a Fibonacci number, as explained in
   the text.
  }
  \label{fig:fig9}
\end{figure}

\begin{figure}[b]
  \centering
  \includegraphics[width=0.9\columnwidth]{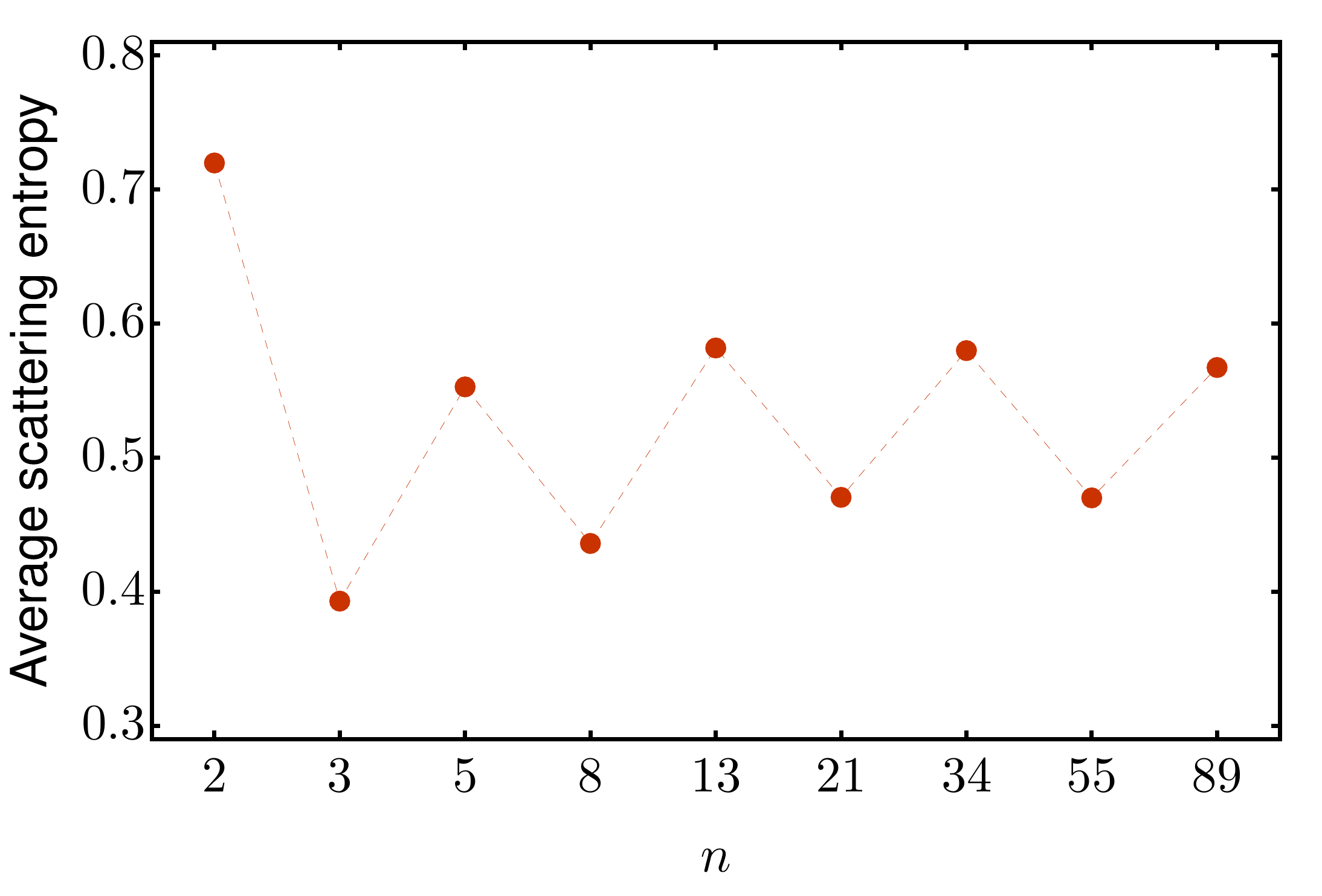}
  \caption{(Color online)
   Average scattering entropy of quantum graphs on the Circle2
   identified by a Fibonacci number.
  }
  \label{fig:fig10}
\end{figure}

Motivated by the possibility of using aperiodic arrangements of
vertices, we then implement the calculation of the ASE.
The results are depicted in Fig. \ref{fig:fig9} on the line and on the
circle, for several Fibonacci numbers.
Notice that the horizontal axis in Fig. \ref{fig:fig9} is out of scale,
to leave room to display several Fibonacci numbers. Moreover, we remind
that the Fibonacci sequence is given by $1,1,2,3,5,8,13,21,\cdots$, with
the property that the sum of any two neighbor  elements gives the next
one, that is, $x_i+x_{i+1}=x_{i+2}$,
with $i=1,2,3,\cdots$.
We also display the results for configurations of the Circle2 type in
Fig. \ref{fig:fig10}.
We compare the results for the red dots in Figs. \ref{fig:fig9} and in
Fig. \ref{fig:fig10} to see that they engender similar qualitative
behavior, although in the Circle2 case the variations are more
significant.

\subsection{Random arrangements}
\label{sec:random}

\begin{figure}[t]
  \centering
  \includegraphics[width=0.9\columnwidth]{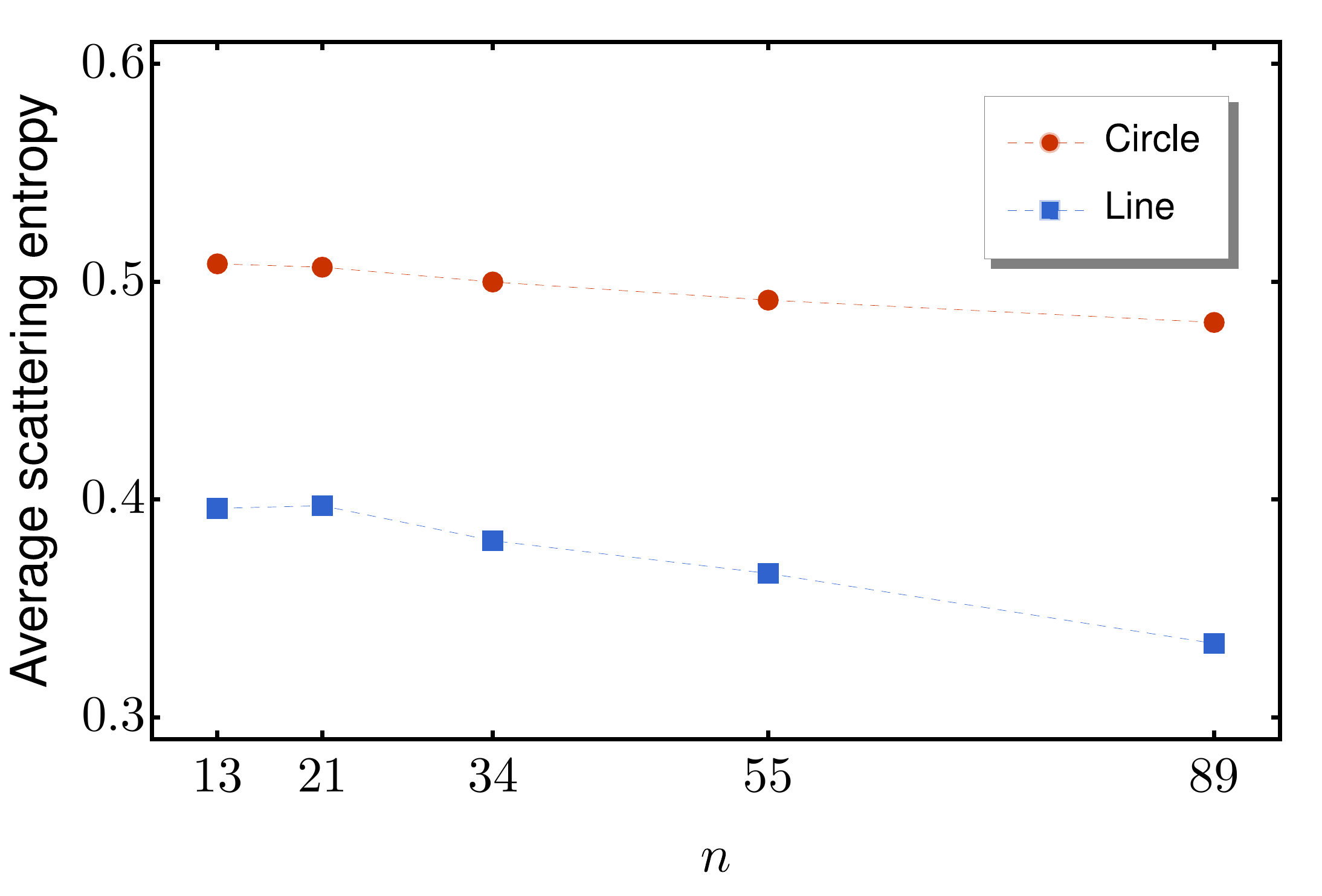}
  \caption{(Color online)
   Average scattering entropy of quantum graphs in which the degree of
   the vertices can be $3$ or $4$, chosen randomly in both the open
   (blue) and closed (red) arrangements of vertices.}
  \label{fig:fig11}
\end{figure}

\begin{figure}[b]
  \centering
  \includegraphics[width=0.9\columnwidth]{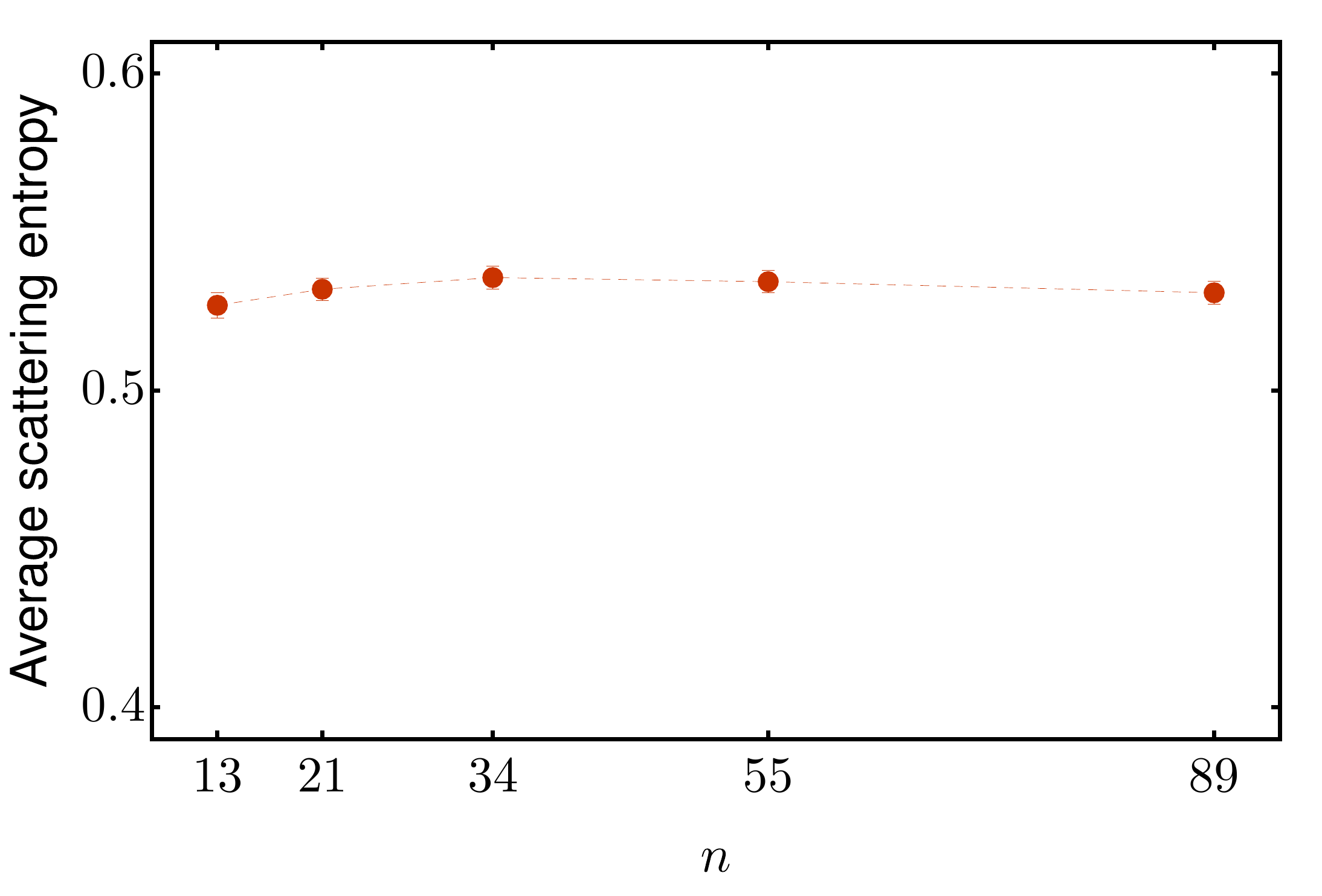}
  \caption{(Color online)
   Average scattering entropy of quantum graphs with $\alpha$ and
   $\beta$ vertices chosen randomly in the Circle2 type of
   arrangements.}
  \label{fig:fig12}
\end{figure}

We now focus on the case where random distributions of
$\alpha$ and $\beta$ vertices are considered.
We also consider the two cases on the line and on the circle,
but now, due to the random nature of the distribution of
  $\alpha$ and $\beta$ vertices, we considered the cases with $13$,
$21$, $34$, $55$ and $89$ vertices.
Evidently, we can choose other values, but the selected Fibonacci
numbers are taken to ease comparison with the results of the previous,
aperiodic arrangements.
To calculate the results displayed in Fig. \ref{fig:fig11}, we
considered  from each ensemble a sample of $100$ different random
choices for the random sequence of $\alpha$ and $\beta$ vertices.
The dots depicted in Fig. \ref{fig:fig11} represent the mean or
expected values, and the corresponding (almost invisible)
bars stand for the standard deviations.
The results show that the ASE on the Circle are
higher than on the line.
We also depict in Fig. \ref{fig:fig12} the case with random arrangements
of $\alpha$ and $\beta$ vertices for configurations of the Circle2 type.

\subsection{Other 2D and 3D results}

\begin{figure}[t]
  \centering
  \includegraphics[width=0.9\columnwidth]{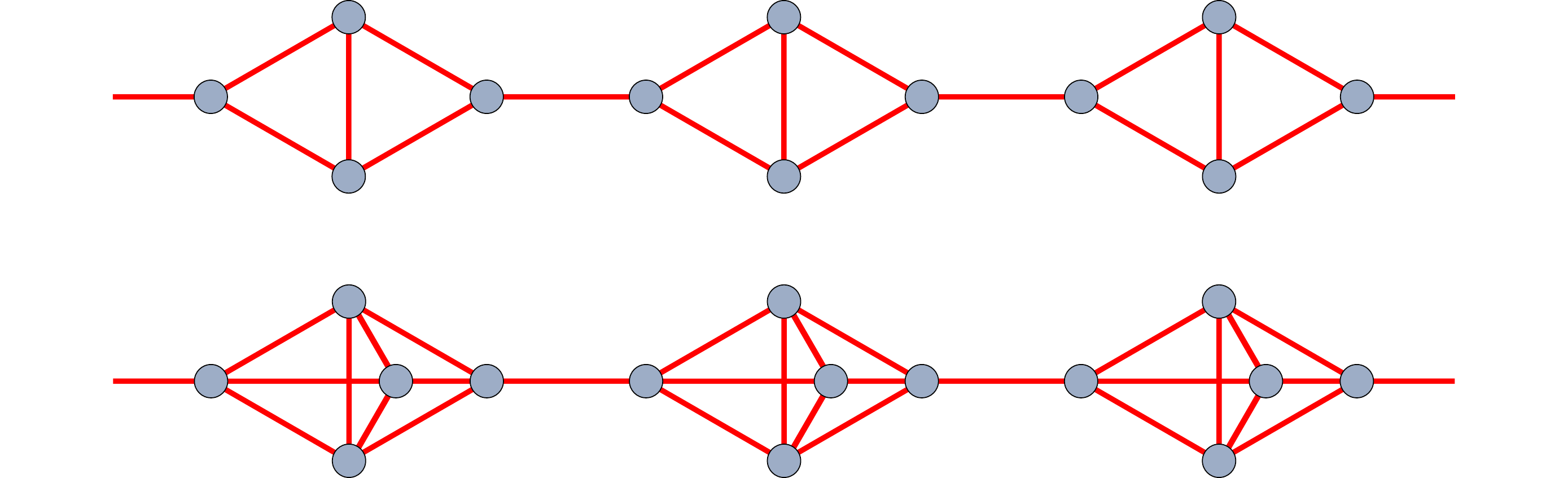}
  \caption{(Color online)
    Two quantum graphs with three planar $\gamma$ structures (top) and
    three spatial $\delta$ structures (bottom), which illustrate the
    general case of periodic structures of the $\gamma_n$ and $\delta_n$
    type, respectively.}
  \label{fig:fig13}
\end{figure}

\begin{figure}[b]
  \centering
  \includegraphics[width=0.9\columnwidth]{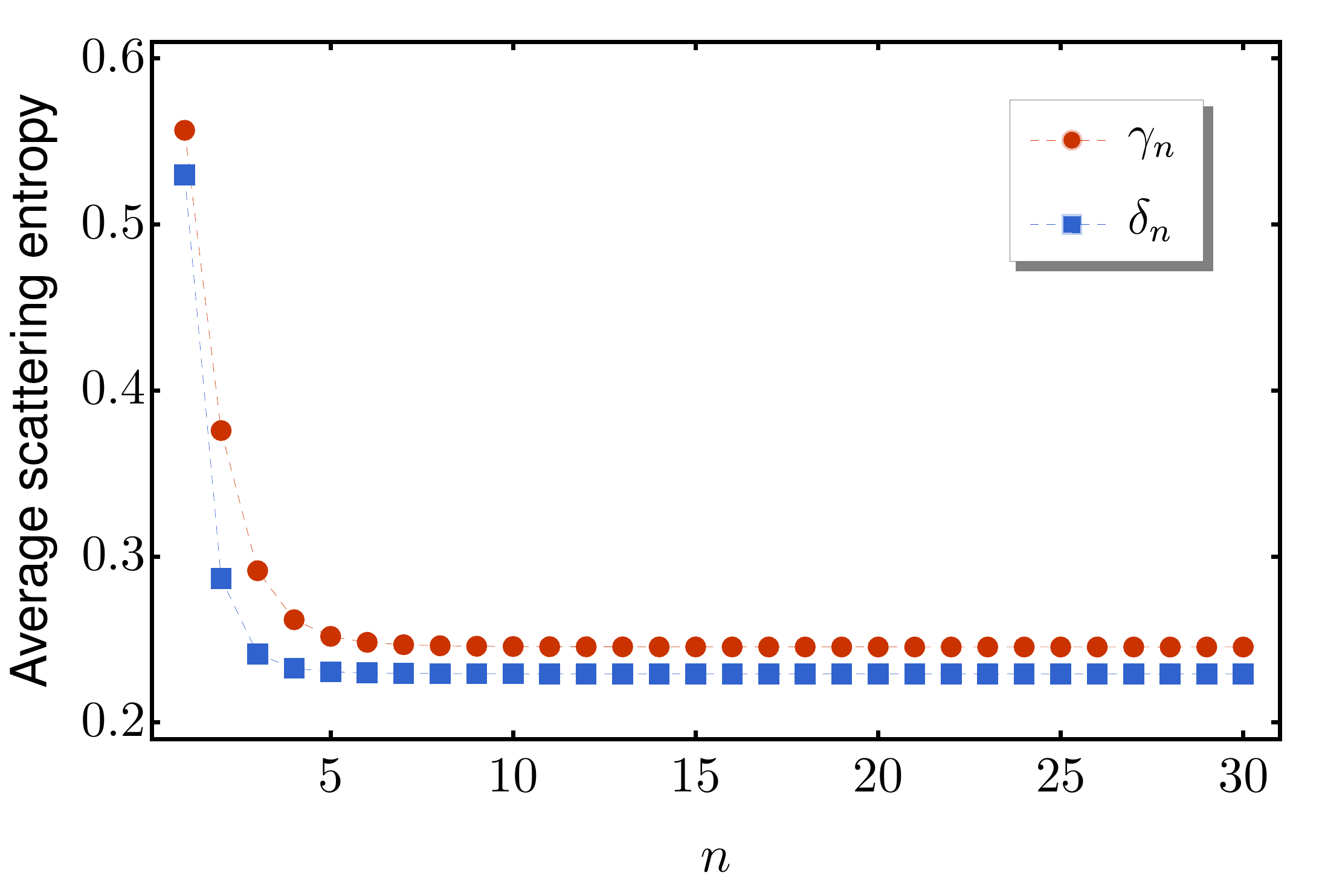}
  \caption{(Color online)
   Average scattering entropy of quantum graphs for the regular
   arrangements of the $\gamma$ structure in 2D and of the $\delta$
   structure in 3D.}
  \label{fig:fig14}
\end{figure}

We can also investigate other planar 2D and spatial 3D arrangements of
QGs.
We first consider the cases having the planar $\gamma$ and the
spatial $\delta$ structures, the first with 4 vertices and 5 edges,
and the second with 5 vertices and 9 edges; see Fig. \ref{fig:fig13}
for an illustration. We notice that in the
2D case, all the vertices have degree 3 and in the 3D case all the
vertices have degree 4.
The $\gamma$ structure may be constructed by fusing two equilateral
triangles, gluing two edges into a single one; the $\delta$ structure
may follow similar procedure, since it can be constructed by fusing two
tetrahedrons, gluing two equilateral triangles into a single one.
We consider periodic dispositions of $\gamma$ and $\delta$ on the line,
and calculate the ASE in the several cases.
The results are displayed in Fig. \ref{fig:fig14} and show similar
qualitative behavior, with the 3D results being always lower than the 2D
ones.

\begin{figure}[t]
  \centering
  \includegraphics[width=0.9\columnwidth]{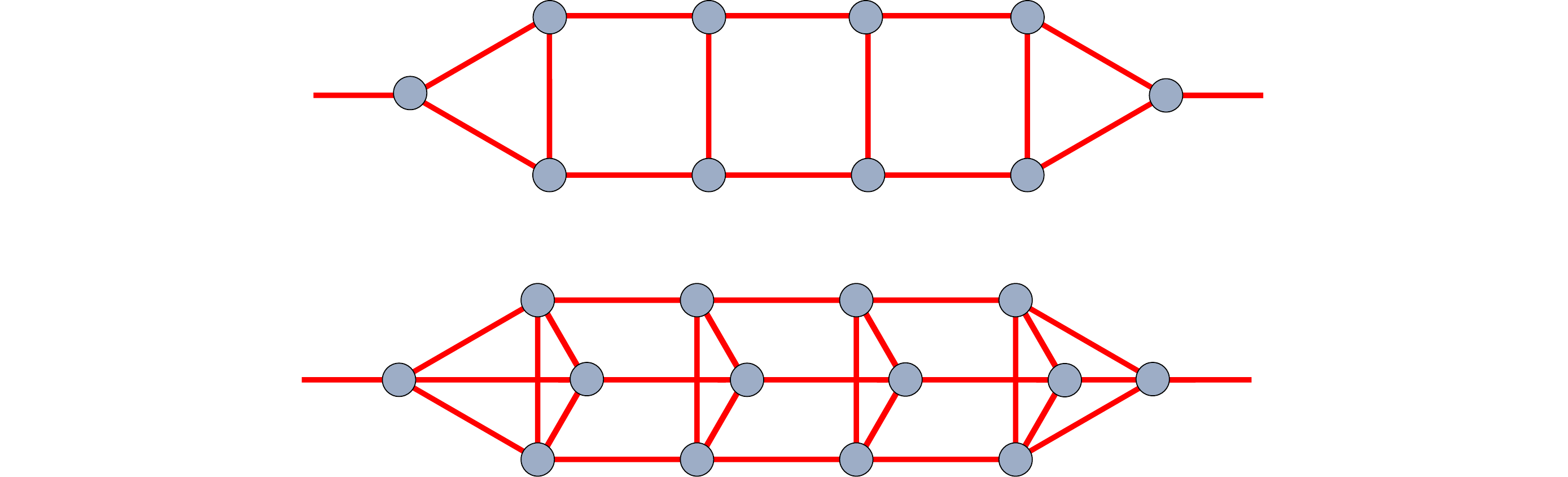}
  \caption{(Color online)
    Two quantum graphs with three squares (top) and three regular
    triangular prisms (bottom), which illustrate the general case of
    periodic structures with squares and prisms, respectively.}
  \label{fig:fig15}
\end{figure}

We can suggest several other possible arrangements of the $\gamma$ and
$\delta$ structures, in particular, the aperiodic and random
dispositions, and this may be implemented following some of the above
investigations.
However, let us now consider two other families of graphs, one composed
of squares in the plane, and the other of regular triangular prisms in
space.
Since we want to have all the vertices with the same degree to
make a fair comparison, in the case of squares, we attach two regular
triangular structures to the regular arrangements of squares, and one
simple external lead to each one of the two triangles, as illustrated in
Fig.  \ref{fig:fig15} with three squares.
In the 3D case, we follow similar steps, but now we consider
arrangements of regular triangular prisms, attaching two tetrahedrons to
their left and right sides and two external leads, as also illustrated
in Fig. \ref{fig:fig15} in the case of three prisms.

\begin{figure}[b]
  \centering
  \includegraphics[width=0.9\columnwidth]{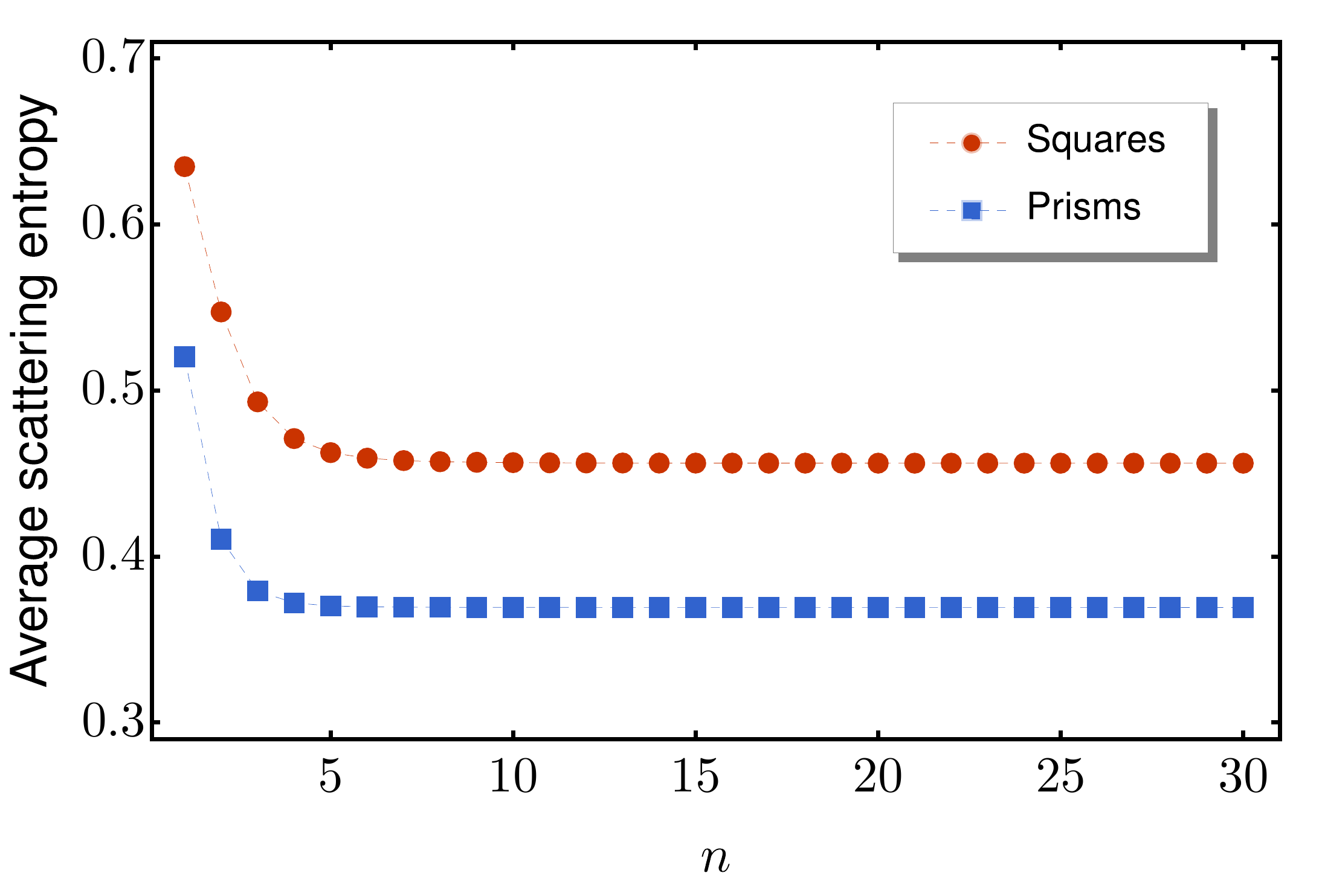}
  \caption{(Color online)
   Average scattering entropy of quantum graphs for the regular
   arrangements of squares in 2D and of triangular prisms in 3D.}
  \label{fig:fig16}
\end{figure}

We see that in the 2D arrangements, all the vertices have degree $3$,
and in the 3D cases, all the vertices have degree $4$.
Recall that we are using regular dispositions of vertices and edges.
We then calculate the ASE for the corresponding QGs and depict the
results in Fig. \ref{fig:fig16} for several squares and prisms,
respectively.
These two families of QGs give interesting results, with the
corresponding ASE being lower for the 3D case.
Also, they follow similar behavior, as in Fig. \ref{fig:fig4}, and
rapidly diminish, saturating to a constant value as one increases the
number of squares and prisms.
The 3D study can be seen as a possibility to use the ASE to investigate
tube-like configurations of current interest in elastic systems and
fluids.
Despite the intrinsic difficulty to study 3D systems, recent
advancements, in particular, in the study of the transition from a
turbulent flow to the case of a coherent flow in three-dimensional
active fluids \cite{S.355.1979.2017} is an interesting motivation.

\section{Conclusion}
\label{sec:end}

In this work we have investigated the average scattering entropy of
quantum graphs in several distinct situations of current interest.
The concept was introduced very recently in
\cite{PRA.103.062208.2021}, and it is based on the Shannon entropy
\cite{Book.Shannon.1963}.
Interestingly, it associates to a given quantum graph a
numerical value, in this sense relating the complicated energy-dependent
expression for the quantum scattering probability of the
quantum graph to a single global quantity.
We have studied cases on the line and on the circle with
several distinct possibilities, with periodic, aperiodic and random
distributions of the vertices.
In particular, we noticed that the ASE results on the circle are
in general higher, when compared to the
corresponding results on the line.
Also, the numerical values seem not to depend on the number of
replications, if there are many replications, but they vary importantly
in the case of small structures, with a small number of replications.

We have studied other families of quantum graphs, some of the
planar type, and others, having spatial configurations.
In the first case, we considered the two possibilities illustrated in
Fig. \ref{fig:fig13}.
The results are displayed in Fig. \ref{fig:fig14}, and they have a
behavior that is qualitatively similar to the case depicted in
Fig. \ref{fig:fig4}.
They also show that the planar arrangements of vertices always give
higher values for the ASE.
In the second case, we investigated a planar family of graphs which is
described by a regular replication of squares, and the spatial family
contains a regular distribution of triangular prisms,
see Fig. \ref{fig:fig15}, and they end up with triangles and tetrahedrons, to ensure that all
vertices have the very same degree: $3$ for the planar case and $4$ for
the spatial case.
The results displayed in Fig. \ref{fig:fig16} follow similar behavior,
with the profiles as the ones depicted in Fig. \ref{fig:fig4}, rapidly
diminishing, saturating to a constant value as one increases the number
of squares and prisms, respectively.

The diversity of results obtained in this work indicates the feasibility
and robustness of the calculations, and they encourage us to further
study the subject, hoping that the ASE may become another tool of
current interest.
We notice, in particular, from the results depicted in
Figs. \ref{fig:fig4}, \ref{fig:fig6}, \ref{fig:fig7}, \ref{fig:fig8}, \ref{fig:fig14} and \ref{fig:fig16}
that the ASE for small structures offers the most large variations, so
we think it appears appropriate to use it to study small molecules, for
instance, the charge transport through single-molecule junctions, which
is quantum mechanical in essence and may provide new tools for the
observation of effects that are not accessible in bulk materials;
see, e.g., the recent review \cite{NRP.1.211.2019} on single-molecule
electronic devices.
In this context, if we think of carbon-based molecules, which are made
of carbon open chains or lines and/or closed rings or circles, since
carbon atoms may form single, double or triple bonds, the study of
quantum transport in small molecules suggests that we consider quantum
multigraphs, in particular the case of two vertices
directly connected by two or three edges.

Other lines of investigation concern the study of the ASE in connection
with thermodynamics, statistical and information quantities, such as
R\'enyi and Tsallis entropies, Fisher information and information and
geometry \cite{Book.Amari.2016}, to quote some interesting
possibilities.
Moreover, in \cite{PRL.111.030602.2013}, for instance, the authors
describe the possibility to transfer energy from a cold system to a hot
system, with the increase in the Shannon entropy of the memory
compensating the decrease in the thermodynamic entropy arising from the
flow of heat against a thermal gradient.
In the same direction, in \cite{NP.11.131.2015} another theoretical
framework for the thermodynamics of information based on stochastic
thermodynamics and fluctuation theorems is presented, where information
can be manipulated at molecular or nanometric scales.
In another direction, in high energy physics in the study of holography
and entanglement entropy, there are many recent works that explore
holographic R\'enyi entropy in conformal field theory
\cite{NC.7.12472.2016}, gravity dual of R\'enyi entropy
\cite{PRL.116.251602.2016}, sensitivity to initial conditions in the
entanglement of local operators, topological lower bound on quantum
chaos \cite{PRL.126.160601.2021} and domain wall topological
entanglement entropy \cite{PRL.126.141602.2021}.
Some of the above issues are currently under consideration, and we hope
to  report on them in the near future.

\section*{Acknowledgments}
This work was partially supported by the Brazilian agencies Conselho
Nacional de Desenvolvimento Cient\'ifico e Te\-cnol\'ogico (CNPq),
Instituto Nacional de Ci\^{e}ncia e Tecnologia de Informa\c{c}\~{a}o
Qu\^{a}ntica (INCT-IQ), and Para\'iba State Research Foundation
(FAPESQ-PB, Grant 0015/2019).
It was also financed by the Co\-or\-dena\c{c}\~{a}o de
Aperfei\c{c}oamento de Pessoal de N\'{i}vel Superior
(CAPES, Finance  Code 001).
FMA and DB also acknowledge CNPq Grants
434134/2018-0 (FMA), 314594/2020-5 (FMA),
303469/2019-6 (DB) and 404913/2018-0 (DB).

\bibliographystyle{model1-num-names}




\end{document}